\documentclass[12pt]{JHEP3}
\usepackage{bm}
\usepackage{amssymb,amsmath,amsthm}
\usepackage{mathrsfs} 
\usepackage{marginnote}
\RequirePackage{graphicx}

\newcommand{\be}{\begin{equation}} 
\newcommand{\ee}{\end{equation}}
\newcommand{\bea}{\begin{eqnarray}}
\newcommand{\eea}{\end{eqnarray}}

\newcommand{\gapp}{\mathrel{\raise.3ex\hbox{$>$}\mkern-14mugo
              \lower0.6ex\hbox{$\sim$}}}
\newcommand{\lapp}{\mathrel{\raise.3ex\hbox{$<$}\mkern-14mu
              \lower0.6ex\hbox{$\sim$}}}

\newcommand\lsim{\lesssim}
\newcommand\gsim{\gtrsim}

\newcommand\vev[1]{{\langle {#1} \rangle}}
\renewcommand\({\left(}
\renewcommand\){\right)}
\renewcommand\[{\left[}
\renewcommand\]{\right]}

\newcommand\eq[1]{Eq.~(\ref{#1})}
\newcommand\eqs[2]{Eqs.~(\ref{#1}) and (\ref{#2})}
\newcommand\eqss[3]{Eqs.~(\ref{#1}), (\ref{#2}), and (\ref{#3})}

\newcommand\eqreff[1]{(\ref{#1})}

\newcommand\pa{\partial}

\newcommand\mpl{M_{\rm P}}

\newcommand{\dlabel}[1]{\label{#1}}

\def\calp{{\cal P}}

\def\calpz{{\calp_\zeta}}

\newcommand\bfk{{\mathbf k}}

\newcommand\bfw{{\mathbf W}}
\newcommand\bfx{{\mathbf x}}

\newcommand\sub[1]{_{\rm #1}}
\newcommand\su[1]{^{\rm #1}}

\newcommand\mone{^{-1}}
\newcommand\mtwo{^{-2}}
\newcommand\mthree{^{-3}}

\newcommand\mfive{^{-5}}
\newcommand\mhalf{^{-1/2}}
\newcommand\half{^{1/2}}

\newcommand\mn{{\mu\nu}}

\newcommand{\fnl}{f\sub{NL}}

\newcommand\bfkp{{{\bfk}'}}

\newcommand{\calpzphi}{\calp_{\zeta_\phi}}
\newcommand{\calpzw}{\calp_{\zeta_W}}

\newcommand{\zetaphi}{\zeta_\phi}

\newcommand{\zetaw}{\zeta_W}

\newcommand{\tend}{t\sub{end}}

\newcommand{\mdf}[1]{#1}

\title {The  statistically anisotropic curvature perturbation  generated by
$f^2(\phi)F_\mn F^\mn$}
\author{David H. Lyth \\
Consortium for Fundamental Physics, Cosmology and Astroparticle Group, 
Department of Physics, Lancaster University, 
Lancaster LA1 4YB, UK \\ 
 E-mail: \email{d.lyth@lancaster.ac.uk}}
\author{Mindaugas Kar\v{c}iauskas \\ 
CAFPE and Departamento de F\'isica Te\'orica y del Cosmos, Universidad
de Granada, Granada-18071, Spain \\ 
University of Helsinki and Helsinki Institute of Physics, P.O. Box 64, FI-00014, 
Helsinki, Finland\\
E-mail: \email{mindaugas.karciauskas@helsinki.fi}}

\abstract{The inflaton might be coupled to a gauge field through
a term $f^2(\phi)F_\mn F^\mn$. If  
$f\propto a\mtwo$ where $a(t)$ is the scale
factor, the perturbation $\delta \bfw$ of the gauge field
 generates a potentially observable
statistically anisotropic contribution
 to the primordial 
curvature perturbation {\em during} slow-roll inflation. 
The  spectrum and bispectrum of this contribution 
have been calculated  using the in-in
formalism of quantum field theory. We  give a simpler and more
complete calculation
using  only the classical perturbations.
The results suggest that either the 
 entire curvature perturbation $\zeta$ (both the statistically isotropic and  
anisotropic parts) is  generated during slow-roll inflation,
or else  it is generated afterwards.}

\preprint{}

\begin{document}

\section{Introduction}

\dlabel{intro}

The observed primordial curvature perturbation $\zeta(\bfx)$ presumably originates in the
early universe, and  a central task of theoretical cosmology
is to determine its origin and the  evolution $\zeta(\bfx,t)$. 
It is usually  supposed to be generated
from the perturbation of one or more scalar fields, making it
 statistically isotropic 
which is consistent with observation.   But it 
may also receive a contribution from one or more vector fields, making it statistically
anisotropic. Observation does allow significant anisotropy and in particular a spectrum
  of the form\footnote
{We use hats to indicate unit vectors}
\be
\calpz(\bfk) = \calpz(k) \[ 1 + g_* (\hat\bfw \cdot \hat \bfk)^2 \]
 \dlabel{gstardef} \ee
with  $|g_*|\lsim 10\mone$ \cite{anisWMAP}. \mdf{The latest results of the \emph{Planck} satellite are consistent with zero $g_*$ at $3\sigma$ level \cite{anisPlanck}.  However, the upper bound on $g_*$ implied by the \emph{Planck} results is not computed yet, though it is expected to be $|g_*|\lsim 10\mtwo$ \cite{anis}.} 

 Most schemes  invoke a  $U(1)$ gauge field $B_\mu$ \mdf{(see e.g. \cite{vecreview} for a review).}
 If the field has the canonical kinetic
term and no coupling to gravity, its 
spectrum increases with wavenumber like $k^2$, 
 and its contribution to the 
 spectrum of $\zeta$ has the same behaviour making it  almost certainly 
  negligible on cosmological
scales \cite{ouranis}. One way of avoiding this may be to keep the canonical kinetic
term but invoke a coupling to gravity
given by $-R B_\mu B^\mu/6$. This apparently  \cite{ouranis,nonmin}
 generates a flat spectrum 
but the model has instability and (possibly) non-linearity \cite{ghost} and it is not known
 \cite{health} whether or not these spoil the prediction. 

In this paper we consider a different scheme, which  invokes a 
gauge kinetic function $f^2(\phi)$ that depends on the slowly
 rolling inflaton field $\phi$. To allow a significant contribution on cosmological scales
one needs  (at least approximately)
 $f\propto a\mtwo$ where $a(t)$ is the scale factor, which may be reasonable in the context
of string theory \cite{p12jcap}.
A well-defined contribution to $\zeta$ is generated during slow-roll inflation 
\cite{beforemarco,marco} and additional
contributions may be generated afterwards \cite{ouranis,p12jcap,jiro}. 
In this paper we give for the first
time a master equation that includes all contributions, but our main focus is on the
contribution generated during slow-roll inflation. Under the assumption 
that the homogeneous part of the 
gauge field dominates the perturbation, it has been calculated 
\cite{beforemarco,marco}
using the in-in formalism of quantum field theory. Using instead
the classical formula for $\zeta$ in terms of the energy
density perturbation, we reproduce that calculation and extend it to the case where
the  homogeneous part is  sub-dominant or even negligible. We also identify the
 assumptions made by the calculation, and show how to perform a more complete
calculation with weaker assumptions. 
We end by discussing the significance of our result.

\section{Action and field equations}

\dlabel{saction}

During inflation we are taking the action to be 
\be
S = \int d^4x \sqrt{-g} \[ \frac12 \mpl^2 R
- \frac12 \pa_\mu \phi \pa^\mu \phi - V(\phi)
 - \frac14 f^2(\phi) F_\mn F^\mn \]
, \dlabel{action} \ee
where $F_\mn\equiv \pa_\mu B_\nu-\pa_\nu B_\mu$  with  $B_\mu$  a 
 gauge field, and $f$ is some function of the inflaton field $\phi$.

Extremizing the action in \eq{action} with respect to fields $\phi$, $B_\mu$ and their derivatives we obtain field equations 
\begin{eqnarray}
\left[\partial_{\mu}+\partial_{\mu}\ln\sqrt{-g}\right]\partial^{\mu}\phi+V'+
\frac{1}{2}ff'F_{\mu\nu}F^{\mu\nu}&=&0; \dlabel{basic1}  \\
\left[\partial_{\mu}+\partial_{\mu}\ln\sqrt{-g}\right]fF^{\mu\nu}&=&0,
\dlabel{basic2}
\end{eqnarray}
where $g\equiv \mathrm{det} (g_{\mu\nu})$, $V'\equiv\partial V/\partial \phi $ and 
$f'\equiv\partial f/\partial\phi$. We make the gauge choice $B_0=0$, 
which fixes the spatial components $B_i(\bfx,t)$ up to a constant.

We are going to assume that the inflationary expansion is nearly
isotropic \mdf{\cite{hair}}, checking later that
this is  justified. Writing
\be
ds^2 = -dt^2 + a^2(t) \delta_{ij} dx^i dx^j
, \ee
we then have
\begin{eqnarray}
\ddot\phi+3H\dot\phi-a^{-2}\nabla^{2}\phi+V' & = & -\frac{1}{2}f f' F_{\mu\nu}F^{\mu\nu}, 
\dlabel{EoM-phi} \\
\ddot B_i+\left(H+2\frac{\dot f}{f}\right)\dot{B}_{i}-a^{-2}\nabla^{2}B_{i} & = & a^{-2}2\frac{\partial_{j}f}{f}\partial_{j}B_{i}, \dlabel{EoM-W}
\end{eqnarray}
where $H\equiv \dot a/a$ and $\nabla^2 \equiv \delta_{ij}\pa^2/\pa x^i\pa x^j$.

\section{Pure slow-roll inflation and the curvature perturbation}

\dlabel{pure}

If the effect of the gauge field is completely negligible
 we have pure slow-roll inflation, described for instance
in \cite{review,book}.
 The last term of \eq{basic1} is absent, and
for   the  unperturbed inflaton $\phi(t)$ we have 
\be
\ddot\phi+ 3H\dot\phi + V'(\phi) = 0
. \ee
The potential is supposed to satisfy  the 
 flatness conditions  $\epsilon\ll 1$ and $|\eta|\ll 1$ where
$\epsilon \equiv \mpl^2(V'/V)^2/2$ and $\eta\equiv \mpl^2 V''/V$. Then, more or less
independently of the initial condition  
\be
3H\dot\phi \simeq -V'
. \dlabel{phidot} \ee
With the flatness conditions this is called the slow-roll approximation, which
we use throughout.
It  gives $|\dot H|/H\ll H$ and $|\dot\epsilon|/H\ll \epsilon$. Except where stated
we take  $H$ and $\epsilon$ to be constant.

We    write $\phi(\bfx,t)=\phi(t) +\delta\phi(\bfx,t)$ and define 
\be
\delta\phi_\bfk(t) \equiv \int d^3x e^{-i\bfk\cdot\bfx} \delta \phi(\bfx,t)
, \ee
and similarly for other quantities.

Let $N(k)$ be the number of $e$-folds of slow-roll inflation after
the epoch of horizon exit $k=aH$. The `cosmological scales' on which $\zeta_\bfk$
is observed have $N(k_0)-15\lsim N(k) \lsim N(k_0)$, where
 $k_0=a_0H_0$ and the subscript on the right hand side denotes the present epoch.
Also, 
\be                                                                         
15 \lsim  N(k_0) \lsim  70
, \dlabel{nbound} \ee
where the upper bound assumes $P\leq \rho$ after inflation, were $P$ is the pressure and $\rho$
the energy density. The lower bound is 
needed         so that cosmological scales leave the horizon. 
Typical cosmologies give a value in the upper third of the range.

Perturbing \eq{basic1} to first order, and ignoring 
 the metric perturbation
(back-reaction) we have
\be
\delta\ddot\phi_\bfk + 3H\delta\dot\phi_\bfk + (k/a)^2 \delta\phi_\bfk= 
-V'' \delta\phi_\bfk 
. \dlabel{phiddot} \ee
Evaluating the metric perturbation to first order, on
 the flat slicing of spacetime
(such that the 3-curvature scalar vanishes) one finds \cite{book,sasmuk}
that it is significant only
after horizon exit, when \eq{phiddot} becomes 
\be
 \delta\ddot\phi_\bfk + 3H\delta\dot\phi_\bfk \simeq 
-  V'' \delta\phi_\bfk - 6H^2\epsilon  \delta\phi_\bfk 
. \dlabel{phiddot3} \ee
The last term  represents the effect of the 
metric perturbation (back-reaction).
{}From \eq{basic1} this  term can be written 
\be
a\mthree \pa_0 \delta(\ln\sqrt{-g}) a^3 \dot\phi \simeq   
3H \delta(\ln\sqrt{-g}) \dot\phi
. \dlabel{backreaction} \ee
Comparing \eqs{phiddot3}{backreaction} 
we find\footnote{We assume $\dot\phi<0$ as is the case for hybrid inflation.}
\be
\delta(\ln\sqrt{-g}) \simeq \frac{-\sqrt{2\epsilon}}\mpl \delta\phi_\bfk
. \dlabel{deltaln} \ee

Regarding $\delta\phi_\bfk(t)$ as an operator its mode function
$\phi(k,t)$ satisfies  \eq{phiddot}. Well before horizon
exit at $k=aH$  \eqs{action}{phidot} describe a free field in flat spacetime.
We  choose
\be
a(t)\phi(k,t) = e^{-ik/aH}/\sqrt{2k}
, \dlabel{phiinit} \ee
and the vacuum state corresponding to the absence of $\phi$ particles. Then
we have
\bea
\vev{ \delta\phi_\bfk \delta\phi_\bfkp } &=&
 (2\pi)^3\delta^3(\bfk+\bfkp)(2\pi^2/k^3) \calp_\phi(k,t), \dlabel{vev} \\
(2\pi^2/k^3)\calp_\phi(k,t) &=&  \phi^2(k,t).
\eea
Also, 
\bea
\vev{
\delta\phi_{\bfk_1} \delta\phi_{\bfk_2} \delta\phi_{\bfk_3} } &=& 0,
\dlabel{phi3} \\
\vev{\delta\phi_{\bfk_1} \delta\phi_{\bfk_2} \delta\phi_{\bfk_3} \delta\phi_{\bfk_4} } 
&=& \vev{\delta\phi_{\bfk_1} \delta\phi_{\bfk_2} } 
\vev{\delta\phi_{\bfk_3} \delta\phi_{\bfk_4} } + \nonumber \\
&+& \vev{\delta\phi_{\bfk_1} \delta\phi_{\bfk_3} }
\vev{\delta\phi_{\bfk_4} \delta\phi_{\bfk_2} } +
\vev{\delta\phi_{\bfk_1} \delta\phi_{\bfk_4} }
\vev{\delta\phi_{\bfk_3} \delta\phi_{\bfk_2} }
 \dlabel{phicorr} , \eea
and similarly for higher products.

Well after  horizon exit ($k\ll aH\ll 1$) 
the phase of $\phi(k,t)$ becomes constant which
means that $\delta\phi$ can be regarded as a classical perturbation
with the above  correlators (gaussian perturbation). 
In the initial regime
\be
|\eta|,\epsilon\ll k/aH \ll 1
, \ee
the left hand  side of \eq{phiddot} is still close to zero, and the mode function
is 
\be
\phi(k,t)\simeq \phi_0(k,t) \equiv \frac{e ^{-ik/aH}}{a\sqrt{2k}}
 \( 1-iaH/k \) \simeq \frac{iH}{k\sqrt{2k}}
. \dlabel{phi0} \ee
This gives a slowly varying perturbation, $H\mone |\delta\dot\phi_\bfk|
\ll |\delta\phi_\bfk|$ with
\be
\calp_\phi(k,t) \simeq  \calp_\phi\su 0\equiv  \( \frac H{2\pi} \)^2 
. \dlabel{pphi0} \ee
The subsequent evolution is given by \eq{phiddot3}. It has two independent solutions
but  the initial slow variation will pick out the solution
\be
 3H\delta\dot\phi_\bfk \simeq 
- \[  V''+ 6H^2\epsilon \] \delta\phi_\bfk 
. \dlabel{phiddot2} \ee

Going to second order in the perturbations of the field and metric one finds
additional correlators of $\delta\phi$ (non-gaussianity) \cite{seery},
starting with the three-point correlator which defines the bispectrum
$B_\phi$:
\be
\vev{\delta\phi_{\bfk_1} \delta\phi_{\bfk_2} \delta\phi_{\bfk_3} }
=(2\pi)^3 \delta^3(\bfk_1+\bfk_2+\bfk_3) B_\phi(\bfk_1,\bfk_2,\bfk_3)
. \dlabel{bphi} \ee

To define the primordial curvature perturbation $\zeta(\bfx,t)$, one
 smoothes  the metric and the energy-momentum tensor
on a super-horizon scale and chooses the comoving
threading and the uniform energy-density slicing of spacetime.
Then 
\be
\zeta(\bfx,t) = \delta \ln(a(\bfx,t)) =\delta\[ \ln a(\bfx,t)/a(t_1) \]
\equiv \delta [N(\bfx,t,t_1)] \dlabel{deltaN}
, \ee
where $a(\bfx,t)$ is the local scale factor such that a comoving volume element
is proportional to $a^3$, and  $a(t_1)$ is its unperturbed
value which can be evaluated at any epoch. As we discuss in Section \ref{just}
it is usually enough to work to first
order in $\zeta$; then one can choose $t_1=t$ to get
\be
\zeta(\bfx,t) = H\delta t(\bfx,t) = -H\delta\rho(\bfx,t)/\dot\rho(t)
, \dlabel{zeta} \ee
where $\delta t$ is the displacement of the uniform-$\rho$ slice from the   
flat slice and $\delta\rho$ is evaluated on the flat slice and
\be
\dot\rho = - 3H(\rho+P)
. \dlabel{econ} \ee

Since $\rho$ is smoothed on a super-horizon scale, \eq{econ} applies at each location,
which implies  that  $\zeta(\bfx,t)$ is constant during any era when $P(\bfx,t)$ 
is a unique function of $\rho(\bfx,t)$.
Galaxy surveys and observation of the
CMB anisotropy detect the value of $\zeta$ at the epoch with temperature
$T\sim 10\mone$\,MeV; the universe is then radiation-dominated ($P=\rho/3$)
giving $\zeta$ a constant value that we  denote   simply by $\zeta(\bfx)$. Its spectrum is nearly scale-independent with \mdf{\cite{planck}}\footnote
{The spectrum and bispectrum of any perturbation are defined as in \eqs{vev}{bphi}.}\mdf{
\bea
\calpz(k) &\simeq&   (5\times 10\mfive)^2   \dlabel{calpzobs} \\  
n(k)-1 &\equiv& d\ln\calpz(k)/d\ln k  \simeq 
-0.0397\pm0.0073
. \dlabel{nobs} \eea
The result for $n(k)$ assumes that it has negligible scale dependence
and tensor fraction, both of which are consistent with observations
\cite{inflation}.}

For the reduced bispectrum defined by
\bea
\fnl(\bfk_1,\bfk_2,\bfk_3) &\equiv& \frac56 \frac {B_\zeta}
{P_\zeta(k_1)P_\zeta(k_2) +P_\zeta(k_2)P_\zeta(k_3) +P_\zeta(k_3)P_\zeta(k_4)}
\dlabel{fnldef} 
 \eea 
where $P_\zeta(k) \equiv (2\pi^2/k^3) \calpz$, 
current observations \mdf{\cite{plngauss} 
give $|\fnl|\lsim 10$. For $\fnl$  to ever be observable we need  
$|\fnl|\gsim 1$.}

During pure slow-roll inflation,
  $\zeta$ has a  time-independent value that we denote
by $\zetaphi$. It is well-approximated by \eq{zeta} with
\be
\rho_\phi  \equiv V(\phi) + \frac12 \dot\phi^2 \simeq V(\phi)
, \ee
leading to the  gaussian perturbation $\zeta=\zetaphi\equiv   
-H\delta\phi/\dot\phi$ and 
\be
\calpzphi(k) =  \frac1{2\epsilon \mpl^2} \( \frac H{2\pi} \)^2
, \dlabel{calpzphidef} \ee
where  $\epsilon$ and $H$ are evaluated at horizon exit. 

Going to higher order in the perturbation of the metric and of $\delta\phi$,
one finds \cite{seery,lms} a non-zero 
$B_\phi$.
Then, using the $\delta N$ formula to go beyond the first-order formula
\eqreff{zeta}, one can use \eqss{vev}{phi3}{phicorr} to find
  $|\fnl|\sim (\calpzphi/\calpz)^2 10\mtwo$ which
 is too small to observe    \cite{seery,lms}.\footnote
{This result was first found in \cite{maldacena}
by a different method.}  

\section{Background (unperturbed) universe}

\dlabel{background}

Recasting \eqs{EoM-phi}{EoM-W}
 in terms of $\mathbf W$ with components
$W_i \equiv f  B_i /a$ and dropping gradient terms, one arrives at equations of motion 
for homogeneous fields $\phi(t)$ and $\bfw (t)$
\begin{equation}
\ddot\phi+3H\dot\phi+V'=\frac{f'}{f}\left|
\dot{\mathbf{W}}+\left(H-\frac{\dot{f}}{f}\right)\mathbf{W}\right|^{2}, \dlabel{hom-phi}
\ee
\be
\ddot{\mathbf{W}}+3H\dot{\mathbf{W}}+\left(2H^{2}-H\frac{\dot{f}}{f}-
\frac{\ddot{f}}{f}\right)\mathbf{W}=0.  \dlabel{hom-W}
\end{equation}

 We  assume   $f\propto a\mtwo$ so that the last
 term of \eq{hom-W} vanishes and  there is a solution with
$\dot\bfw=0$. The other solution is decaying, $\bfw\propto a\mthree$, and   
 corresponds to 
$B_i=$\,constant. With the action \eqreff{action} this 
can be set to zero
by a  gauge transformation and has no  physical effect. We therefore take
$\bfw(t)$ to be constant.
(For hybrid inflation with the waterfall coupling to $\bfw$ both solutions are
physical \cite{p12jcap} but we  assume that the decaying solution is anyhow negligible.)
It would make no essential difference if  we allow $f\propto a^\alpha$
with $\alpha$ slightly different from $-2$, or more generally if we just assume
\be
3H\dot \bfw \simeq - \left(2H^{2}-H\frac{\dot{f}}{f}-\frac{\ddot{f}}{f}
\right)\mathbf{W},\qquad H\mone|\dot\bfw| \ll W\equiv |\bfw|
. \ee

We assume that the right hand side of \eq{hom-phi} is small enough that the approximation
$3H\dot\phi\simeq -V'$ still applies. Then
$f'/f \simeq  2/\sqrt{2\epsilon} \mpl$ and treating the right hand side as a first
order perturbation we have
\be
3H\dot \phi \simeq - V' +  \frac{18 H^2 W^2}{\sqrt{2\epsilon} \mpl}\simeq -V'
. \dlabel{homphi2} \ee

The energy density of the gauge field  is \cite{fF2curv} 
$\rho_{W}(\bfx,t)=-f^{2}F_{\mu\nu}F^{\mu\nu}/4$. Smoothed on a super-horizon scale
this gives
\be
\rho_{W}=\frac92 H^2 W^2
\dlabel{rhow}. \ee
Using this result for the unperturbed energy density, we see that  our
assumption that the right hand side of
 \eq{homphi2} is dominated by the first term is consistent if
\be
\frac{2\rho_W}{\epsilon\rho} \equiv \frac{3 W^2}{\epsilon\mpl^2} \ll 1 ,
\dlabel{rhowcon}\ee
which we assume.

\section{Perturbed universe}

\dlabel{perturbed}

Evaluating  \eqs{EoM-phi}{EoM-W} to first order in $\delta\phi$ and 
dropping the right hand side of \eq{EoM-W} we get 
\bea
\delta\ddot\phi + 3H\delta\dot\phi - \nabla^2\delta \phi
&=& -V'' \delta \phi +  \frac2{\sqrt{2\epsilon}\mpl} \delta \[ \left|
\dot{\mathbf{W}}_\bfx+3H \mathbf{W}_\bfx \right|^{2} \] - \nonumber \\
& - &\frac{8H\mathbf{W}_\bfx}{\sqrt{2\epsilon}\mpl}  \left(
\dot{\mathbf{W}}_\bfx+3H\mathbf{W}_\bfx\right)
\frac{\delta\dot\phi}{\dot\phi}  \dlabel{a9} \\
\delta\ddot \bfw+3H \delta\dot\bfw - \nabla^2 \delta \bfw
&=& \frac{ 2\bfw_\bfx }{ \sqrt{2\epsilon}\mpl } \[ \delta\ddot\phi
-3H\delta\dot\phi - \nabla^2 \delta\phi \]
 \dlabel{a10} \\
\bfw_\bfx &\equiv& \bfw(\bfx,t) \equiv \bfw + \delta\bfw(\bfx,t) ,
\eea
where $\bfw$ is the unperturbed value.

This ignores the metric perturbation (back-reaction). 
Since slow-roll inflation is supposed to be a good approximation, the effect of the 
 metric perturbation in  \eqs{basic1}{basic2} is significant 
only well after horizon exit, and 
is then given by  \eq{deltaln}. Including it, \eqs{a9}{a10} become well after 
horizon exit
\bea
\delta\ddot\phi + 3H\delta\dot\phi 
&=& -\(V''+6H^2\epsilon \) \delta \phi \nonumber + \\
&+&  \frac2{\sqrt{2\epsilon}\mpl} \delta \[ \left|
\dot{\mathbf{W}}_\bfx+3H \mathbf{W}_\bfx\right|^{2} \] 
-\frac{8H\mathbf{W}_\bfx}{\sqrt{2\epsilon}\mpl}  \left(
\dot{\mathbf{W}}_\bfx+3H\mathbf{W}_\bfx\right)
\frac{\delta\dot\phi}{\dot\phi}  \dlabel{a93} \\
\delta\ddot \bfw+3H \delta\dot\bfw 
&=& \frac{ 2\bfw_\bfx }{ \sqrt{2\epsilon}\mpl } \[ \delta\ddot\phi
-3H\delta\dot\phi + \frac32
H  \dot\epsilon \delta\phi \]
, \dlabel{a103} 
\eea
where in \eq{a93} the  back-reaction $-6H^2\epsilon\delta\phi$ is the same as 
in \eq{phiddot2}. 

Let us first set the  right hand sides of \eqs{a9}{a10} to zero.
We saw in Section \ref{homphi2} how the vacuum fluctuation of $\delta\phi$
is then converted at horizon exit
to a nearly gaussian classical perturbation, and the same thing happens
to the  vacuum fluctuation of $\bfw_\bfk$  \cite{ouranis}.\footnote
{In contrast with the case for scalar field perturbations \cite{seery},
the non-gaussianity of $\delta\bfw$ has yet to be evaluated, but we will assume
that it still has a negligible effect.} 
Its  left- and right-handed components have the same 
mode function $W(k,t)=\phi_0(k,t)$ in \eq{phi0}, which gives
well  after horizon exit 
\be
 \vev{ \delta W_\bfk^i \delta W_\bfkp^j } 
= (2\pi)^3 \(\delta^{ij} - \hat k^i \hat k^j \)
\delta^3(\bfk+\bfkp) (2\pi^2/k^3)(H/2\pi)^2   \dlabel{wspec}
. \ee

Now we consider the effect of the right hand sides of \eqs{a9}{a10}.
In the regime  $k\gg aH$, spacetime curvature is negligible and we deal with field
theory in flat spacetime. 
The theory involves massless gauge bosons and the nearly massless
inflaton particles ($|V''|\ll (k/a)^2$ 
which propagate as  nearly free particles (perturbative regime), justifying
the initial condition   \eqreff{phiinit}.

To discuss the subsequent evolution, let us  take $\delta\phi(\bfx,t)$
and $\delta\bfw(\bfx,t)$ to include only  modes with $k$ in some small interval,
so that there is a well-defined
epoch of horizon exit. 
 In the (still quantum) regime $k\sim aH$,
 $H$ is the  only relevant scale and  we have typical magnitudes
\bea
\delta\phi \sim\delta \bfw\sim H,\qquad 
\delta\dot\phi \sim\delta \dot\bfw\sim H^2,\\ 
\delta\ddot\phi \sim\delta \ddot\bfw\sim \nabla^2 \delta\phi
\sim \nabla^2 \delta\bfw \sim H^3
. \eea
Using these with \eq{rhowcon}, we see that
 the right hand sides of
\eqs{a9}{a10} are much smaller than $H^3$, 
and hence have only a small effect.\footnote
{The findings of this  and the previous paragraph remain valid when the last term of
\eq{a10} is included, and that term is negligible in the super-horizon regime
that we are about to discuss.}
 It is therefore reasonable to  assume that the perturbations become
 classical  soon after  horizon 
exit,  with \eqs{pphi0}{wspec} initially a good approximation giving 
typical values $|\delta\phi| \sim|\delta \bfw|\sim H$.

The evolution of classical perturbations is described by 
\eqs{a93}{a103} with $k=0$. 
One of their solutions is 
\begin{eqnarray}
3H\delta\dot\phi &\simeq&  - \( V'' + 6H^2\epsilon \) \delta\phi+
\frac{18 H^2}{\sqrt{2\epsilon}\mpl}   \delta\( W^2(\bfx,t) \) \\
&\equiv& - \( V'' + 6H^2\epsilon \) \delta\phi+
\frac{18 H^2}{\sqrt{2\epsilon}\mpl}   
\( 2\bfw\cdot\delta\bfw + \delta\bfw\cdot\delta\bfw \)
\dlabel{a92}  \\
\delta\dot\bfw  &\simeq& -\frac{2\bfw}{\sqrt{2\epsilon}\mpl}
 \delta\dot\phi , \dlabel{a102} 
\end{eqnarray}
which implies $|\delta\dot\phi|\ll H|\delta\phi|$
and $|\delta\dot\bfw|\ll H|\delta\bfw|$.
The  self-consistency of this solution can be
checked by inserting it into the right hand sides of \eqs{a92}{a102}.
It is presumably  picked out by the initial condition, 
just as in the case of slow-roll inflation.

Since all of the quantities appearing in the second term of \eq{a92}
 vary slowly on the Hubble timescale,
we will  take them all to be constant, giving
\bea
\delta\phi_\bfk(\tend) &=& \delta\phi_\bfk\su 0(\tend) + 
\frac{6N(k)}{\sqrt{2\epsilon}\mpl}   \( \delta(W^2)\)_\bfk 
, \eea
where $\delta\phi_bfk\su 0(\tend)$ is the slow-roll result.

\section{Contribution of $\delta\bfw$ to the  curvature perturbation}

\dlabel{contribution}

Now we calculate $\zeta$ 
at the end of slow-roll inflation, using \eq{zeta}. Smoothed on a super-horizon
scale the energy density is $\rho=\rho_\phi+\rho_W$ 
\bea
\delta\rho(\bfx,\tend) &\simeq& V'\delta\phi(\bfx,\tend)  + \frac92 H^2\delta(W^2) \\
&\simeq& V' \delta\phi_0(\bfx,\tend) + 18N(k)H^2 \delta(W^2)
+ \frac92 H^2\delta(W^2) \\
&\simeq& V' \delta\phi_0(\bfx,\tend) + 18N(k)H^2 \delta(W^2)
. \eea
This gives
 $\zeta_\bfk(\tend)=\zeta^\phi_\bfk +\zeta\su W_\bfk$,  where  
\bea
\zeta^W_\bfk &=& \frac12 C(k) \delta(W^2)_\bfk \nonumber \\
&=& C(k) \[ \bfw\cdot\delta\bfw_\bfk + \frac12 \[ \(\delta\bfw\)^2\]_\bfk \]
\dlabel{zetawofc},
\eea
with 
\be
C(k) = \frac{6N(k)}{\epsilon\mpl^2} \dlabel{cofk} 
. \ee

We can now  calculate the spectrum and bispectrum of
$\zeta_W$ using \eq{wspec} and the analogues of \eqs{phi3}{phicorr}. 
The result depends on the value of the  unperturbed field $\bfw$,
and in particular on its magnitude $W$.

\subsection{Working in a finite box}

An  unperturbed quantity  is the zero mode of its Fourier expansion, and at this
point we need to remember that within the cosmological context that  
expansion has to be done within 
 some   box of finite coordinate size $L$  \cite{mybox}. 
Assuming nearly exponential inflation, the box
 leaves the horizon $N_L(k)\equiv
\ln(kL)$ $e$-folds
before the scale $k$. 
To    avoid making unverifiable assumptions about
an era that will never be constrained by observation (and, as we will see,
also to simplify the calculation) one should  take  $k_0L$ to be 
as small as is consistent with the requirement that the
periodic boundary condition implied 
by the use of the Fourier expansion have  a negligible effect  (minimal box). 
 Demanding say $1\%$ accuracy in the
calculation, it should be enough to choose 
$k_0L\sim 100$ corresponding to $\ln(k_0L) \sim 5$. 
After choosing $L$ one writes  for a given quantity
$g(\bfx,t)= g(t) + \delta g(\bfx,t)$, where the `unperturbed' value
$g(t)$ is the average of $g$ within the box. 

Expectation values like \eqreff{vev}  are in general  defined
with respect to an ensemble of universes, 
one of which is the observed universe.
But  under the usual assumption that perturbations originate 
as a vacuum fluctuation
the translation invariance of the vacuum makes them translation invariant.
As a result the expectation values can be  
 defined as spatial averages within a single realisation of the ensemble.
Thus the $\vev{(\delta g)^2}$ 
can be defined as the spatial average within the box,
 $\vev{\delta g(\bfx+{\bf X}) \delta g(\bfx) }$ can be defined as the average
with respect to ${\bf X}$ and so on \cite{book}.

Keeping only classical modes, \eqs{pphi0}{wspec} give
\bea
\vev{(\delta\phi)^2)} &=& \int^k_L \calp_\phi(k) dk/k =\ln(k_0L) (H/2\pi)^2  \\
\vev{ |\delta \bfw|^2 } &\simeq& 2 \ln(k_0L)(H/2\pi)^2 \dlabel{deltawsq1}
. \eea
For the minimal box size this corresponds to typical values
$|\delta\phi|\sim |\delta\bfw|\sim H$.

\subsection{The value of  $W$}

As was discussed in \cite{mybox}, there
 are two possible viewpoints about the  magnitude of an  unperturbed
field like $\bfw$.\footnote
{These  apply to  
 any non-inflaton field that acquires a perturbation from its vacuum
fluctuation. 
The inflaton field is an exception because the inflation model and the cosmology
determines its value $N(k_0)$ before the end of inflation.
We ignore the issue of anthropic selection, assuming that statistical anisotropy
(like non-gaussianity \cite{book})
is neither favoured or disfavoured in that respect.}
One is to regard the magnitude
 $W$ as  a free parameter, that one gets to choose just like 
the masses and couplings appearing in the action. 
The other viewpoint  is to 
 estimate the likely value of $W$,  assuming that we are at a typical
location within  a box whose   size $M$ is very much bigger than the size $L$ of the
minimal box within which the calculations are done. 

Adopting the second  viewpoint one has to make an assumption about the evolution 
of the universe long before the observable universe leaves the horizon. 
The usual assumption is that there is almost
exponential  inflation,  beginning 
$N_M$ $e$-folds before the observable
universe leaves the horizon with $N_M$ fairly large. One can then estimate 
the likely value with suitable assumptions about the relevant physics during 
those $e$-folds. In our case, let us assume that 
the dependence $f\propto a\mtwo$ 
continues to (at least approximately) hold during those $e$-folds.
Then,
after smoothing $\bfw$ on the scale $L$ \eq{wspec} gives for the average within
the exponentially inflated patch
\be
\vev{ |\delta \bfw|^2 } \simeq 2 \ln(M/L)(H/2\pi)^2 \simeq 2 N_M(k) (H/2\pi)^2
, \dlabel{deltawsq} \ee
Therefore, the expected value of $W^2$, if our location is typical, is
\be
W^2 = W_M^2 + \ln(M/L) (H/2\pi)^2 > \ln(M/L)(H/2\pi)^2
= N_M(k_0) (H/2\pi)^2
, \dlabel{nmeq} \ee
where $W_M$ is the average within the inflated patch. 
We conclude that  if we occupy a typical location within an 
inflated patch that left the horizon many $e$-folds before the observable
universe, then  $W\gg H$. 

\subsection{The case $W\gg H$}

If $W\gg H$, the second term in the square bracket of \eq{zetawofc}
can be treated as a first-order perturbation. Then, 
assuming that $\zetaw$ gives the only contribution to the 
anisotropy of $\calpz$,  Eq.~(39) of \cite{ouranis}  gives 
\be
\calpz(\bfk) = \calpz(k)  \[ 1 + g_*(k) \(\hat\bfw\cdot \hat \bfk\)^2 \] \dlabel{ourcalpz}
, \ee
with 
\bea
g_*(k) &=&  - \calpzw(k)/\calpz(k)  \dlabel{gofc} \\
 &=& -\frac{ C^2(k) W^2 }{ \calpz(k)} \( \frac H{2\pi} \)^2 \dlabel{calpzw2} \\
&=& -48N^2(k) \frac{\rho_W}{\epsilon\rho} \frac{\calpzphi}{\calpz(k)}. \dlabel{ourg}
\eea 
Here, $\calpzphi$ is given by \eq{calpzphidef}, and is independent of $k$ since we
are taking $H$ and $\epsilon$ to have negligible time-dependence.

The anisotropy is of the form \eqreff{gstardef} but with a strongly scale-dependent $g_*(k)\propto N(k)$.
This  has not been compared with observation but the constraint is
presumably similar to the scale-independent case, 
 \mdf{discussed after \eqreff{gstardef}}.
Another constraint  comes from the strong scale-dependence of
 $\calpzw(k)$, which gives a contribution
\mdf{$-g_*(k)$
to the spectral index $n(k)$}. Using \eq{nobs}, this requires barring a cancellation \mdf{$|g_*(k_0)|\lsim 0.04$}.
In any case, one certainly needs $|g_*(k_0)|\ll 1$, which with 
 \eq{ourg} is  stronger than \eq{rhowcon}, justifying the latter.
Our assumption that the anisotropy of the expansion has a negligible effect
is also justified, because that anisotropy is 
 presumably only of  order $(\rho_W/\rho)$, which presumably gives 
a contribution $(\rho_W/\rho)(\calpzphi/\calpz(k))$ to $g_*$ which is
 smaller than the one that we have calculated.

Using  $\calpz(k_0)=(5\times 10\mfive)^2$,
\be
g_*(k_0) \simeq 
 -1.3\times 10\mthree \( \frac{N(k)}{60} \)^2 \( \frac{\calpzphi}{\calpz(k_0)}
\)^2 \( \frac WH \)^2  \dlabel{geqn}
. \ee
If we were to assume $\calpzphi=\calpz$ as in \cite{marco}, 
the observational bound \mdf{$|g_*|\lsim 10^{-2}$} leads to two conclusions as
those authors notice. First,  our assumption
$W\gg H$ would be only marginally allowed. Also, from
 \eq{nmeq}, one sees that $N_M$ to be very large if we occupy a typical
location.

Now we calculate the contribution of
$\zeta_W$ to  the bispectrum $B_\zeta$ on cosmological scales.
To simplify the calculation we  set $N(k)=N(k_0)$.
 Taking $\calpz(k)$ to be scale-independent and defining
\be
\frac65\fnl(\bfk_1,\bfk_2,\bfk_3) \equiv \frac1{4\pi^4}\frac {B_\zeta}
{\calpz^2}\frac{\prod k_i^3}{
\sum  k_i^3}
, \ee 
Eq.~(41)  of  \cite{ouranis} gives
\be
\fnl(\bfk_1,\bfk_2,\bfk_3) = \fnl  \( 1 +
f\sub{ani}(\bfk_1,\bfk_2,\bfk_3) \)
, \dlabel{fnlanis} \ee
where the constant prefactor is given by
\bea
\fnl&=&  \frac56 \frac{g_*^2(k_0)}{C(k_0)W^2} \dlabel{fnl1x} \\
&=& -10 N(k_0)  g_*(k_0) (\calpzphi/\calpz), \dlabel{fnl2x}
\eea
and 
\be
f\su{ani} = \frac{
 - (\hat \bfw \cdot \hat \bfk_1)^2
-(\hat \bfw \cdot \hat \bfk_2)^2 + (\hat \bfk_1 \cdot \hat \bfk_2)
(\hat \bfw \cdot \hat \bfk_1)(\hat \bfw \cdot \hat \bfk_2)  }{
\sum k_i^3/ k_3^3  }
+ \mbox{ 2 perms.}
. \dlabel{fnlanis2} \ee
This  expression has not yet been confronted with data, but
the bound is presumably similar to the one that  takes $g_*(k)$ to be
constant, currently \mdf{$|\fnl|\lsim 10$}.

\subsection{The case $W \lsim  H$}

Consider now the case $W=0$. Since there is no preferred direction,
$\zeta_W$ is statisticaly isotropic. Using Eq.~(40)  of \cite{ouranis} we have
\bea
\frac{\calp_{\zeta_W}}{\calpz}  & =& \frac{\calp_{\zeta_W}\su{loop} }{\calpz}
\equiv \frac43 C^2(k) \( \frac H{2\pi} \)^4 \ln(kL)/\calpz
\dlabel{calpz1}  \\
&=& 3N^2(k)\calpz  \ln(kL) \(\frac{ \calpzphi(k)}{\calpz} \)^2 \ll 1
. \dlabel{calpz2} \eea
Using \eq{wspec} and the analogues of \eqs{phi3}{phicorr} we find  
\bea
\fnl&=&f_{NL}\su{loop} \equiv \frac23 C^3 \( \frac H{2\pi} \)^6 
\frac{ \ln(kL) }{ (\calpz)^2 }
\[
  1 + 
\( \frac{ \(\hat \bfk_1\cdot \hat \bfk_2\)^2 }{
\sum_i k_i^3/k_3^3 } + \mbox{c.p.}
\) \]  \dlabel{fnl1} \\
&=&  \frac{3\half}4   \(\frac{\calp_{\zeta_W}}{\calpz} \)^{3/2} 
 \ln\mtwo(kL) \calpz\mhalf
\[
  1 + 
\( \frac{ \(\hat \bfk_1\cdot \hat \bfk_2\)^2 }{
\sum_i k_i^3/k_3^3 } + \mbox{c.p.}
\) \]
\dlabel{fnl2}  \\
&=& \frac94 N^3 \calpz \( \frac {\calpzphi}{\calpz} \)^3 \ln(kL) 
\[
  1 + 
\( \frac{ \(\hat \bfk_1\cdot \hat \bfk_2\)^2 }{
\sum_i k_i^3/k_3^3 } + \mbox{c.p.}
\)\] \ll 1
\eea
In the general case, $\calp_{\zeta_W}$ and $\fnl$ are the sum of the `tree' and
`loop' contributions.

\subsection{How to make the calculation more accurate}

\dlabel{just}

The calculation that we have presented is sufficiently accurate, unless and
until  statistical anisotropy is observed. If that does happen a more accurate
calculation may be required. 

Such a calculation should in principle
begin with the generation of the classical
perturbations $\delta\phi$ and $\delta\bfw$ from the vacuum, along the lines
of \cite{lms}.
 It may in principle generate significant time-dependence and
correlation for these quantities, when they first become classical.
But  the analysis of Section \ref{pure} suggests
that it will instead confirm  the  result obtained there;
nearly time-independent and uncorrelated perturbations with the spectra
$(H/2\pi)^2$ defined in \eqss{vev}{pphi0}{wspec}.

After the perturbations become classical, their evolution 
is given by  \eqs{a93}{a103}, which can
be solved numerical to determine the correlators of the perturbations
at the end of inflation. That allows one to determine $\delta\rho$,
and working to first order in  $\zeta$ we can use \eq{zeta}
to determine $\calp_{\zeta_W}$ and $f\sub{NL}$.

Instead of working with the perturbations, 
one can use the $\delta N$ formula \cite{ouranis}
\bea
\zeta(\bfx,t) &=& \zetaphi(\bfx,t) + \zetaw(\bfx,t) _+
+ \sum_i \frac12 N_{\phi\phi}(t) [ \delta\phi_*(\bfx) ]^2
+ \frac12 N_{\phi i} \delta\phi_*(\bfx) \delta W_i^*(\bfx)\quad\quad  
\dlabel{deltanours}\
 \\
\zetaphi(\bfx,t) &\equiv&   N_\phi(t) \delta\phi_*(\bfx), \\
\zetaw(\bfx,t) &\equiv&  \sum_i N_i(t) \delta W_i^*(\bfx) + \frac12 \sum_{ij}
N_{ij}(t) \delta W_i^*(\bfx) \delta W_j^*(\bfx) 
. \eea
Here $N$ is defined by \eq{deltaN} with $t_*$ the epoch of horizon exit for
the scale of interest, and the
subscripts on $N$ denote partial derivatives evaluated at the unperturbed
point  in the field space.
But as we now argue,  the perturbative approach of the previous
paragraph is expected to be adequate, i.e. the first-order 
\eq{zeta} is expected to be adequate.\mdf{\footnote{This conclusion is 
also supported by ref. \cite{hassanNew}. The authors of this work 
calculated $g_*$ and $\fnl$ using the $\delta N$ formalism. Their 
results agree with our computed values of $g_*$ and tree level $\fnl$ 
up to an overall sign (compare their eq. (4.14) and our eq. (6.18)).}}

The  validity of \eq{zeta} for $\zetaphi$
is a standard result but we need to justify its use for $\zeta_W$.
The second order correction is presumably of order  $\zeta_W^2$, corresponding to
a tiny fractional correction of order $\zeta_W$ that can certainly be ignored
for the evaluation of $\calp_{\zeta_W}$. To see whether it can be
ignored for $\fnl$, let us first pretend that $\bfw$ is a scalar field.
Then, setting $C(k)$ to a constant and assuming that the first term
of \eq{zetawofc} dominates, \eqs{zetawofc}{fnl1x} give\footnote
{If instead the second term dominates, it is solely responsible for
$\fnl$ and  the first-order formula will certainly be adequate.}
\be
\zeta_\sigma(\bfx) = \zeta\sub g(\bfx)
+ \frac35 f\sub{NL} \( \frac{\calpz}{\calp_{\zeta_\sigma}} \)^2
\zeta\sub g^2(\bfx) 
 ,\ee
where $\zeta\sub g\equiv C W \delta W$.
The first-order formula will therefore be adequate unless
\be
f\sub{NL} \lsim \( \frac{\calp_{\zeta_\sigma}}{\calpz} \)^2
. \ee
Since we need $|\fnl|\gsim 1$ for it to be observable,
we conclude that 
the first-order formula will be adequate for the evaluation
of $\fnl$ unless $|\fnl|\sim 1$ {\em and} $\zeta_W$
is the dominant contribution to $\zeta$. Except for the second proviso
this is a standard result, that was first recognised in the context of
the curvaton scenario \cite{beforelr,lr}.

Keeping the vector nature of $\bfw$, \eqs{zetawofc}{fnl1x} give
\bea
\zeta_\sigma(\bfx) &=& \zeta\sub g(\bfx)
+ \frac35 f\sub{NL} \( \frac{\calpz}{\calp_{\zeta_\sigma}} \)^2
\tilde\zeta\sub g^2(\bfx) \\
\zeta\sub g(\bfx) &=& C\bfw\cdot \delta\bfw(\bfx)) \\
\tilde \zeta\sub g(\bfx)  &=& C W |\delta \bfw(\bfk)|  
 \eea
At a typical location, $\tilde \zeta\sub g\sim \zeta\sub g$, and the
previous conclusion about the validity of the first-order formula still
applies.

\section{Conclusion}

\dlabel{conc}

Working exclusive with the classical perturbation of $\bfw$, we  have 
presented a
rather complete   calculation of the contribution to $\zeta_W$ that is
generated during slow-roll inflation. In the regime $W\gg H$ we
 have reproduced the result of
\cite{beforemarco,marco} for the spectrum (\eq{ourg}) and of \cite{marco}
for the bispectrum (\eqs{fnlanis}{fnl2x}, discussing for the first time
the assumptions that are needed to obtain it.

The field $\bfw(\bfx,t)$  may survive after the end of 
slow-roll inflation, and generate further contributions to $\zeta$.
The effect of these is to change $C(k,t)$, so that 
 \eqs{gofc}{fnl1x} as well as \eqss{calpz1}{fnl1}{fnl2}  
still hold but with $C$ the final value.

A contribution $\zeta\sub{end}$ may be generated during the 
waterfall that ends hybrid
inflation \mdf{\cite{p12jcap,jiro}}. It can be calculated from the 
`end-of-inflation' formula,
provided that the waterfall is sufficiently brief which requires 
$H\lsim 10^{-9}\mpl (\calp_{\zeta\sub{end}} /\calpz )\half$.\footnote
{This result is an obvious extension of the one that was derived in
  \cite{p11} with the assumption 
$\calp_{\zeta\sub{end}} =\calpz $. It follows from the fact that
the duration of the waterfall cannot be much less than $1/m$.} The result 
is \mdf{\cite{p12jcap}}
\be
C\sub{end} = -\frac{h^2}{\sqrt{2\epsilon}\mpl mg},
\ee
\mdf{where $m$ is a bare mass of the waterfall field, and the inflaton 
and  vector field coupling constants to the waterfall are denoted by $g$ 
and $h$ respectively.\footnote{See eqs. (3.3) and (3.4) of 
ref. \cite{p12jcap} for precise definitions of these quantities.}}

If the gauge symmetry is spontaneously broken after inflation, the \mdf{vector
curvaton mechanism may generate another contribution after inflation 
\cite{ouranis,kostas}.}
With the simplest  assumption that $\bfw$ is nearly time-independent until
it begins to oscillate 
\be
C\sub{curv} = \frac{\rho_W\su{dec}}{\rho\su{dec}}
\frac1{3 W^2}
, \ee
where `dec' denotes the epoch just before the decay of $\bfw$.

Including both these contributions we have
\be
C = C\sub{sr} + C\sub{end} + C\sub{curv}
, \ee
where $C\sub{sr}=6 N(k)/\epsilon \mpl^2$ is the contribution generated during
slow-roll inflation.
The extra contributions allow  the non-gaussianity to  be observable 
even if $W\ll H$.

We close with an important comment, stemming  from \eq{geqn}.
On the reasonable assumption that there were some large number
$N_M$ of $e$-folds of inflation before the observable Universe left the
horizon, one expects $W/H\sim N_M\gg 1$. But then \eq{geqn} is compatible
with the observational bound on $g_*$ only if $\calpzphi\ll\calpz$;
in other words, if the observed curvature perturbation is mostly
generated {\em after} slow-roll inflation by for instance the 
end-of-inflation or curvaton mechanism.
Our results therefore   suggest a dichotomy
regarding the generation of  observable statistical anisotropy of the curvature
perturbation. 
Either the entire curvature perturbation $\zeta$ (both the isotropic and the 
statistically anisotropic part) is likely to be generated during slow-roll inflation,
or it is likely to be generated afterwards.

\section{Acknowledgments}
DHL 
 acknowledges support from the Lancaster-Manchester-Sheffield Consortium for
Fundamental Physics under STFC grant ST/J00418/1, and from
 UNILHC23792, European Research and Training Network (RTN) grant. 
MK is supported by the grants CPAN (CSD2007-00042) and MICINN (FIS2010-17395)
also by the Academy of Finland grant 131454.


\end{document}